\documentclass[letterpaper, 10 pt, conference]{ieeeconf}  

\usepackage{amsmath,amssymb,amsthm}
\usepackage{graphicx}                       
\usepackage{epstopdf}
\usepackage{siunitx}                        
\usepackage{fancyhdr}                       
\usepackage{balance}                        
\usepackage[hidelinks]{hyperref}            

\graphicspath{{../}{./figures/}}
\UseRawInputEncoding

\newtheorem{proposition}{Proposition}

\usepackage{booktabs}
\usepackage{threeparttable}
\usepackage{multirow}
\usepackage{url}
\usepackage{algorithm}
\usepackage[noend]{algpseudocode}
\usepackage{makecell}
\usepackage{multirow}
\usepackage{cleveref}
\crefname{proposition}{Proposition}{Propositions}
\Crefname{proposition}{Proposition}{Propositions}

\usepackage{tabularray}
\usepackage{tabu}
\usepackage{booktabs}
\usepackage{stfloats}
\usepackage{cite}
\usepackage{cuted}

\usepackage{xcolor,soul,framed} 
\colorlet{shadecolor}{yellow}
\hypersetup{
colorlinks=true,
linkcolor=blue,
anchorcolor=red,
citecolor=blue
}

\IEEEoverridecommandlockouts                              

\title{\LARGE \bf
Direct Data-driven Predictive Control: A Computationally Efficient Alternative to DeePC for Eco-driving in Mixed Traffic Flows
}

\author{Dongjun Li$^{1,2}$, Haoxuan Dong$^{2}$, Liangcai Xu$^{2}$, and Ziyou Song$^{1,2*}$
\thanks{$^{1}$D. Li, and Z. Song is with the Department of Electrical Engineering and Computer Science, University of
Michigan, Ann Arbor, 48109, MI, USA, and also with the Department of Mechanical Engineering, National University of Singapore, 117575 Singapore, Singapore (email: {\tt\small dongjli@umich.edu; ziyou@umich.edu})}
\thanks{$^{2}$H. Dong and L. Xu are with the Department of Mechanical Engineering, National University of Singapore, 117575 Singapore, Singapore (email:
        {\tt\small donghaox@foxmail.com; liangcai@u.nus.edu}) }%
}

\begin{document}

\maketitle
\thispagestyle{empty}
\pagestyle{empty}

\begin{abstract}
Improving energy efficiency in the transportation sector is critical for achieving sustainable mobility, with eco-driving emerging as a key strategy. However, implementing effective eco-driving for connected and automated vehicles (CAVs) in mixed traffic presents a significant control challenge due to the heterogeneous, uncertain behavior of human-driven vehicles (HDVs). Data-enabled Predictive Control (DeePC) offers a promising model-free approach but is often hindered by a high computational burden, limiting its real-time feasibility. This paper introduces a novel Direct Data-driven Predictive Control (D3PC) framework to address this limitation. By reformulating the data-driven prediction mechanism, the D3PC significantly reduces computational complexity, making its computation time nearly invariant to historical data size. This computational efficiency directly enables the formulation of a sophisticated eco-driving controller that can solve the complex energy optimization problem in real time, even within diverse and stochastic mixed-traffic environments. Comprehensive simulations demonstrate that the D3PC is orders of magnitude faster than existing DeePC-based methods while achieving superior energy efficiency. Specifically, it reduces total platoon energy consumption by up to 10.71\% compared to rule-based cruise control baselines and 3.80\% compared to the original DeePC, confirming its effectiveness for real-time, energy-efficient control.
\end{abstract}

\section{INTRODUCTION}
The pursuit of sustainable mobility is critically dependent on greater energy efficiency within the transportation sector, where eco-driving has emerged as a primary method for reducing fuel consumption and emissions \cite{wang2023adaptive,lu2024safe}. The advent of Connected and Automated Vehicles (CAVs) presents a transformative opportunity to systematically implement optimal eco-driving strategies, leveraging their advanced sensing, communication, and computational capabilities \cite{matin2022impacts}. However, the widespread adoption of CAVs will be gradual, leading to a prolonged period of mixed traffic flows where CAVs must coexist and interact with human-driven vehicles (HDVs)  \cite{yao2023analysis}.

This mixed-traffic environment constitutes a complex human-in-the-loop cyber-physical system \cite{li2024physics}. The primary challenge stems from the inherent uncertainties associated with HDVs. Human driving behavior is stochastic, nonlinear, and diverse, making it exceedingly difficult to predict or model with high fidelity \cite{li2023human}. These unpredictable dynamics complicate the design of CAV controllers that can effectively smooth traffic flow and maximize energy savings, demanding advanced control strategies that can operate robustly amidst these uncertainties.

To address this challenge, various control methodologies have been developed. Traditional Model Predictive Control (MPC) offers robust performance but relies heavily on accurate system models, which are difficult to identify for unpredictable HDVs. Model-free approaches like Reinforcement Learning can learn adaptive policies but often suffer from high computational burdens and lack formal safety guarantees \cite{lu2024safe}. More recently, Data-enabled Predictive Control (DeePC) has emerged as a promising alternative, using raw input-output data to bypass explicit model identification \cite{coulson2019data}. Despite its advantages, the standard DeePC framework suffers from a significant drawback: its reliance on large Hankel matrices results in a high computational load, hindering its feasibility for real-time applications.

Recognizing this limitation, recent research has focused on improving the computational tractability of DeePC. One common approach involves dimensionality reduction techniques, such as Singular Value Decomposition (SVD), to compress the data matrix and reduce the number of decision variables \cite{zhang2023dimension}. More advanced strategies leverage machine learning to create computationally cheaper approximations of the DeePC problem. For example, some methods learn a compact system representation or an approximate scoring function to significantly reduce online computation time \cite{zhou2024learning}. Other efforts include developing distributed DeePC algorithms for large-scale systems or employing faster optimization solvers \cite{wang2023distributed}. However, while these methods provide valuable improvements, they often act as incremental enhancements or approximations built upon the original, computationally intensive DeePC paradigm. This motivates the search for a fundamentally more direct and efficient formulation for data-driven control.

To overcome these limitations, this paper introduces a novel Direct Data-driven Predictive Control (D3PC) paradigm. This method reformulates the data-driven control problem to fundamentally reduce its computational complexity. The main contributions of this work are twofold: (1) we propose the D3PC framework, a direct data-driven predictive control method that bypasses system identification and significantly improves computational efficiency through reformulating the prediction mechanism compared to DeePC and subspace predictive control \cite{favoreel1999spc}; (2) we implement and validate this method for the eco-driving control of CAVs in mixed traffic flows, demonstrating its effectiveness and practicality for energy-aware control applications. Our approach paves the way for a more efficient and directly applicable data-driven control solution for complex, real-world systems.

\section{SYSTEM MODELING}\label{model}
This section briefly introduces the linearized mixed traffic flow model, the intelligent driver model (IDM), and the energy consumption model for electric vehicles. 

\subsection{Mixed Traffic Flow Model}
\begin{figure}[!htpb]
    \centering
    \includegraphics[width=1\linewidth]{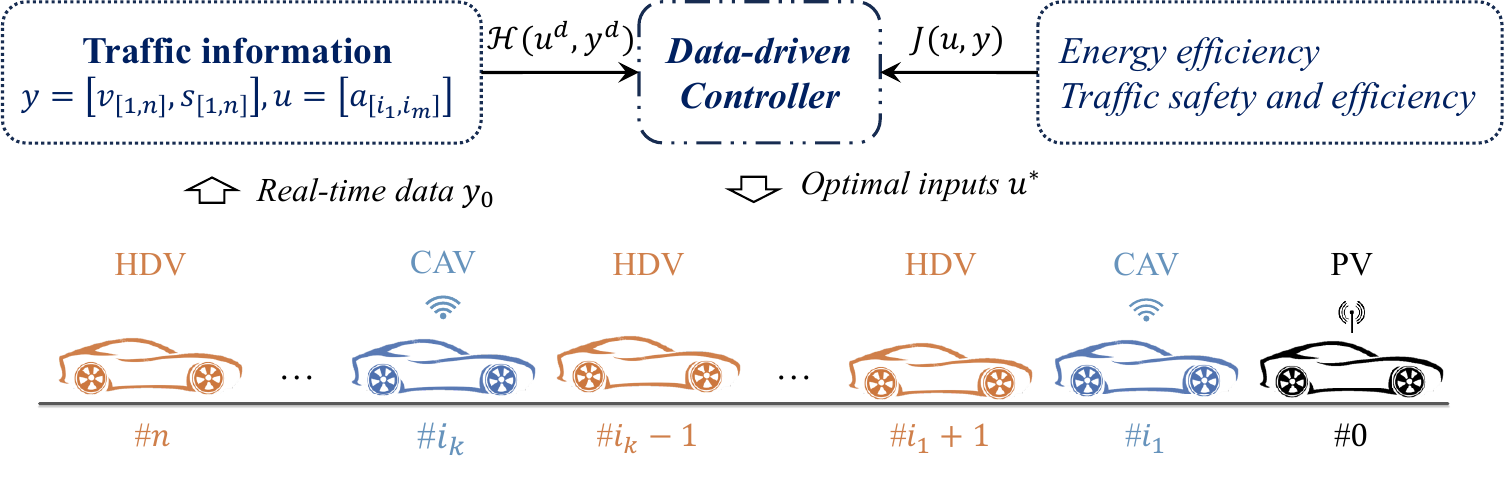}
    \caption{Data-driven control schematic of mixed traffic flows.}
    \label{fig:trafficf_flow}
\end{figure}

Considering a single-lane mixed traffic flow with $n+1$ vehicles indexed as $n \in \Omega =\{1, 2, \ldots, n\}$, which comprising one preceding vehicle (PV) indexed as $0$, $m$ CAVs indexed as $S=\{1,2,\ldots,m\}$ and $n-m$ heterogeneous HDVs indexed as $\Omega \setminus S$, as shown in Figure \ref{fig:trafficf_flow}. Car-following models with different parameters are employed to describe the intricate driving behavior of HDVs. CAVs in mixed traffic flow serve as the control objects. Hence, the car-following dynamics of CAVs and HDVs can be written in the form
\begin{equation}
    \begin{aligned}
        \dot{s}_{i}(t) & = v_{i-1}(t) - v_{i}(t), \qquad \qquad i \in \Omega,  \\
        \dot{v}_{i}(t) & =
        \left\{
            \begin{aligned}
                & F\left(s_{i}(t),\dot{s}_{i}(t),v_{i}(t)\right), && i \in \Omega \setminus S, \\
                & u_{i}(t),                                       && i \in S,
            \end{aligned}
        \right.
    \end{aligned}
\end{equation}
where $s_{i}$ denotes the inter-vehicle spacing between vehicle $i$ and its preceding vehicle $i-1$; $v_{i}$ denotes the velocity of vehicle $i$; $F(\cdot)$ is a nonlinear function, representing car-following models like IDM \cite{treiber2000congested}, optimal velocity model (OVM), or their variants.

The mixed traffic flow dynamics can be linearized around the equilibrium point $(s^{*},v^{*})$. Hence, the spacing and velocity perturbations can be defined as 
\begin{equation}
    \label{eq_equilibrium}
    \left\{
    \begin{aligned}
        & \widetilde{v}_{i}(t) = v_{i}(t) - v^*  && v^{*} = \frac{1}{\Delta{T}}  \int_{t-\Delta T}^{t_{k}} v_{0}(t)dt \\ 
        & \widetilde{s}_{i}(t) = s_{i}(t) - s^*  && s^{*} = F^{-1}\left(\dot{v}_{i}= \dot{s}_{i}=0, v_{i} = v^{*} \right)
    \end{aligned}
    \right.
\end{equation}
where $\Delta{T}$ denotes the past time interval, particularly, $v^{*}$ is the average velocity over the past $\Delta{T}$; $v_{0}$ denotes velocity of the PV.  With the state variables denoted by $x(t)=\left[\widetilde{s}_{1}(t),\widetilde{v}_{1}(t),\ldots,\widetilde{s}_{i}(t),\widetilde{v}_{i}(t),\ldots,\widetilde{s}_{n}(t),\widetilde{v}_{n}(t)\right]^\top$, and CAVs' accelerations as the control inputs, the linear state-space model of mixed traffic flow is defined as:
\begin{equation}
    \label{eq_mixedTraffic}
    \left\{
    \begin{aligned}
        & \dot{x}(t) = Ax(t) + Bu(t) + H\varepsilon(t), \\
        & y(t) = Cx(t),
    \end{aligned}
    \right.
\end{equation}
where $\varepsilon(t) = \widetilde{v}_{0}= v_{0}(t) - v^{*} $, $A \in \mathbb{R}^{2n\times2n}, B \in \mathbb{R}^{2n\times m}, C \in \mathbb{R}^{2n\times 2n}$, and $H \in \mathbb{R}^{2n\times 1}$. 
For more details of Eq. (\ref{eq_mixedTraffic}), please refer to \cite{li2025eco-driving}.

\subsection{Intelligent Driver Model}
The IDM is a widely used car-following model that generates realistic acceleration profiles. Its dynamics are described by:
\begin{equation}
    \centering
    \label{eq_IDM}
    \begin{aligned}
        \dot{v} & = F\left(s,\dot{s},v\right) = F\left(s,\Delta{v},v\right) \\
        & = a\left( 1- \left( \frac{v}{v_{d}} \right)^{\delta} - \left( \frac{s_{\mathrm{st}} + vT_{\mathrm{gap}} - \frac{v\Delta{v}}{2\sqrt{ab}}}{s} \right)^{2} \right),
    \end{aligned}
\end{equation}
where $a/b$ denote the maximum acceleration and deceleration; $v_{d}$ is the desired velocity; $\delta$ is acceleration exponent; $\Delta v =\dot{s}$ is the velocity difference to its preceding vehicle; $s_{\mathrm{st}}$ is the minimum inter-vehicle spacing; $T_{\mathrm{gap}}$ is the time headway.

To capture the diversity of human driver behaviors, the time headway parameter $T_{\mathrm{gap}}$ is selected based on parameter calibration results obtained from field-test vehicle trajectory data provided by the NGSIM project \cite{li2024physics}. Following the calibration procedure described in \cite{wang2023adaptive}, the IDM is tuned using the car-following data of different human drivers in the NGSIM dataset. 

\begin{figure}[!ht]
        \centering
	\includegraphics[width=0.85\linewidth]{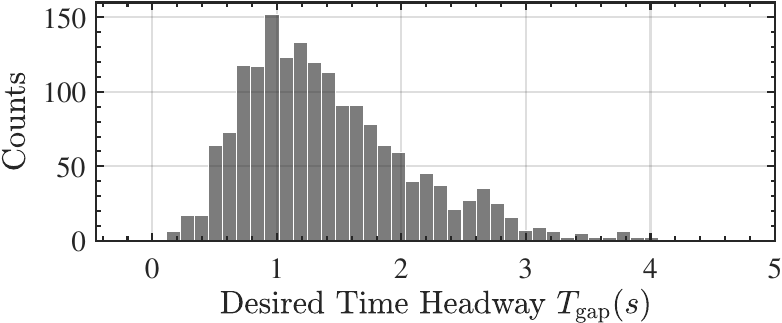}
	\caption{Distribution of desired time headway $T_{\mathrm{gap}}$.} 
	\label{fig_Distribution}
\end{figure}

To strike a balance between optimization performance and computational efficiency, five IDM parameters ($a_{\max}, \delta, s_{\mathrm{st}}, b, v_{d}$) are held constant \cite{wang2023adaptive}. The time headway $T_{\mathrm{gap}}$ is thus treated as the primary parameter representing driver heterogeneity, as illustrated in Figure \ref{fig_Distribution}. 

\subsection{Energy Consumption Model}
We selected electric vehicles as the focus of our research. An approximate and differentiable energy consumption model is employed, which is well-established and maintains accuracy with minimal loss \cite{li2025eco-driving,lu2024safe}. This model is expressed as a polynomial function,
\begin{equation}
	\label{eq_polynomial}
	P_{\mathrm{tot}}(t) = \sum_{i=0}^{3} \sum_{j=0}^{2}p_{ij}v^{i}(t)a^{j}(t),
\end{equation}
where $p_{ij}$ are calibrated parameters detailed in Table \ref{tab:ECModelparams}, $v(t)$ denotes the vehicle velocity, and $a(t)$ denotes the acceleration. To ensure consistency across comparisons, this same energy consumption model is applied to both CAVs and HDVs.
\begin{table}[!htbp]
	\caption{Energy consumption model parameters}
	\begin{center}
		\begin{tabular}{cccccc}
		\toprule
        Parameter & $p_{00}$         & $p_{01}$           & $p_{02}$         & $p_{10}$          & $p_{11}$ \\   
        Value     & $\mathrm{110.3}$ & $\mathrm{1213}$    & $\mathrm{2911}$  & $\mathrm{422.9}$  & $\mathrm{2484}$\\ \hline
        Parameter & $p_{12}$         & $p_{20}$           & $p_{21}$         & $p_{30}$          &  \\   
        Value     & $\mathrm{25.19}$ & $\mathrm{-0.0279}$ & $\mathrm{1.374}$ & $\mathrm{0.3557}$ &  \\
        \bottomrule
        \end{tabular}
	\end{center}
	\label{tab:ECModelparams}
\end{table}

\section{METHODOLOGIES}
\subsection{Preliminaries}
The starting point of DeePC is the behavioral systems theory and Willems’ fundamental lemma \cite{willems2005note}, which state that sufficiently rich input--output data can characterize all trajectories of a discrete-time LTI system, without the need for explicit model identification.  

Consider an input--output trajectory of length $T$,
\[
u^d_{[1,T]} = \mathrm{col}(u_1,\dots,u_T), \qquad
y^d_{[1,T]} = \mathrm{col}(y_1,\dots,y_T),
\]
and construct Hankel matrices of depth $L = T_{\mathrm{ini}} + N$. By partitioning these into past and future components, we obtain
\[
\begin{bmatrix} U_p \\ U_f \end{bmatrix} = \mathcal H_L(u^d_{[1,T]}), 
\qquad
\begin{bmatrix} Y_p \\ Y_f \end{bmatrix} = \mathcal H_L(y^d_{[1,T]}),
\]
where $(U_p,Y_p)$ represent past data of length $T_{\mathrm{ini}}$ and $(U_f,Y_f)$ represent future data of length $N$.

If the collected input $u^d_{[1,T]}$ is persistently exciting of order $L+n$ (with $n$ an upper bound on the system order), Willems’ fundamental lemma ensures that every valid trajectory of the system can be expressed as a linear combination of the columns of these Hankel matrices \cite{willems2005note}. Specifically, for given past inputs and outputs, $u_{\mathrm{ini}} = \mathrm{col}(u_{k-T_{\mathrm{ini}}},\dots,u_{k-1})$, $y_{\mathrm{ini}} = \mathrm{col}(y_{k-T_{\mathrm{ini}}},\dots,y_{k-1})$, and a future input sequence $u = \mathrm{col}(u_k,\dots,u_{k+N-1})$, the corresponding output sequence $y = \mathrm{col}(y_k,\dots,y_{k+N-1})$ will satisfy the following the fundamental lemma:
\begin{equation}
\label{eq:data-eq}
    \begin{bmatrix}
        U_p \\ Y_p \\ U_f \\ Y_f
    \end{bmatrix} g =
    \begin{bmatrix}
        u_{\mathrm{ini}} \\ y_{\mathrm{ini}} \\ u \\ y
    \end{bmatrix}.
\end{equation}
When $T_{\mathrm{ini}}$ is chosen no smaller than the lag of the system, the future outputs $y$ are uniquely determined by $(u_{\mathrm{ini}},y_{\mathrm{ini}},u)$.  

Based on this data-driven representation, DeePC formulates a predictive control problem where the optimization variables are the future input and output trajectories. A typical formulation can be given as follows:
\begin{equation}
    \centering
    \begin{aligned}
        \underset{u,y}{\min} \quad J  & = \sum_{k=0}^{N-1} \left( \Vert y_k - y_{\mathrm{ref},k}\Vert_Q^2 + \Vert u_k\Vert_R^2 \right), \\
        \mathrm{s.t.} \quad & \eqref{eq:data-eq}, \\
        & u_k \in \mathcal{U}, y_k \in \mathcal{Y}.
    \end{aligned}
\end{equation}
In this way, DeePC preserves the structure of MPC, while replacing the explicit model with a purely data-driven behavioral representation \cite{coulson2019data}.

\subsection{Direct Data-driven Model}
Referring to the study in \cite{breschi2023data}, the DeePC formulation can be split into a two-stage scheme: fitting the initial conditions and optimizing the future performance, while guaranteeing constraint satisfaction. Thus, Eq. \eqref{eq:data-eq} can be rewritten as follows:
\begin{equation}
    \centering
    \begin{bmatrix}
        Z_0 \\
        U_f \\
        Y_f 
    \end{bmatrix}g=
    \begin{bmatrix}
        z_{\mathrm{ini}} \\
        u \\
        y
    \end{bmatrix},
    \label{eq:two_stage_data_eq}
\end{equation}
where $Z_{0} = [U_p^{\top},Y_p^{\top}]^{\top}$ and $z_{\mathrm{ini}} = [u_{\mathrm{ini}}^{\top},y_{\mathrm{ini}}^{\top}]^{\top}$, indicating any past joint input and output Hankel matrix and trajectory. 

Particularly, the equality condition $Z_0g = z_{\mathrm{ini}}$ represents the initial condition fitting stage, while the remained equality condition and the cost function represent the future optimization stage. Here, we further exploit and derive this two-stage scheme, thereby proposing the direct data-driven model (D3M), which transforms the implicit fundamental lemma of DeePC into an explicit, direct input--output model that is analogous to a state-space representation. 

We begin with the two-stage scheme of the DeePC for a one-step-ahead prediction. By setting the depth of the future Hankel matrix as $N=1$, Eq. \eqref{eq:two_stage_data_eq} becomes: 
\begin{equation}
    \begin{bmatrix}
        Z_{0} \\
        U   \\
        Y
    \end{bmatrix} g = 
    \begin{bmatrix}
        {z_{\mathrm{ini}}} \\
        {u}_k \\
        {y}_{k+1}
    \end{bmatrix},
    \label{eq_DeePC_D3M}
\end{equation}
where $Z_0$ contains past input-output data, $U$ contains the current input ${u}_k$, and $Y$ contains the future output ${y}_{k+1}$. The vector ${z_{\mathrm{ini}}}$ represents the known real-time past input--output trajectory, and ${g}$ is the latent variable that ensures consistency.

From Eq.~\eqref{eq_DeePC_D3M}, we can derive an explicit data-driven model. The top two blocks of the equation define the relationship between the known data (initial conditions and given input signals) and the latent variable:
\begin{equation}
    \begin{bmatrix}
        Z_{0} \\
        U
    \end{bmatrix} g = 
    \begin{bmatrix}
        {z_{\mathrm{ini}}} \\
        {u}_{k}
    \end{bmatrix}.
\end{equation}

\begin{proposition}\label{proposition1}
    For deterministic LTI systems, Willems' fundamental lemma asserts that the future outputs $y$ are uniquely determined by the given trajectory $(u_{\mathrm{ini}}, y_{\mathrm{ini}}, u)$. Consequently, the latent variable $g$ does not influence this determinism for LTI systems.
\end{proposition}
Building on \cref{proposition1}, the latent variable $g$ can be selected arbitrarily without affecting the system outputs. To obtain a unique representation in practice, the least-norm solution is adopted via the Moore--Penrose pseudo-inverse, which yields:
\begin{equation}
    {g} = 
    \begin{bmatrix}
        Z_{0} \\
        U
    \end{bmatrix}^{\dagger}
    \begin{bmatrix}
        {z_{\mathrm{ini}}} \\
        {u}_{k}
    \end{bmatrix}.
\end{equation}
The bottom block of the original equation, i.e., ${y}_{k+1} = Yg$, defines the predicted output. By substituting the expression for ${g}$, we obtain the formal definition of the D3M for one-step-ahead prediction:
\begin{equation}
    {y}_{k+1} = Y \left( \begin{bmatrix}
        Z_{0} \\
        U
    \end{bmatrix}^{\dagger}
    \begin{bmatrix}
        {z_{\mathrm{ini}}} \\
        {u}_{k}
    \end{bmatrix} \right).
    \label{eq_D3M}
\end{equation}

To reveal the model's structure, we can expand the pseudo-inverse term. Let $\Lambda = Z_{0}^{\top}Z_{0} + U^{\top}U$. The pseudo-inverse can be represented as follows:
\begin{equation}
    \begin{aligned}
        \begin{bmatrix}
            Z_{0} \\
            U
        \end{bmatrix}^{\dagger}
        & =
        \left(\begin{bmatrix}
            Z_{0}^{\top} & U^{\top}
        \end{bmatrix}
        \begin{bmatrix}
            Z_{0} \\
            U
        \end{bmatrix}\right)^{-1}
        \begin{bmatrix}
            Z_{0}^{\top} & U^{\top}
        \end{bmatrix} \\
        & = 
        \underset{\Lambda^{\dagger}}{\underbrace{\left(Z_{0}^{\top}Z_{0} + U^{\top}U\right)^{-1}}}
        \begin{bmatrix}
            Z_{0}^{\top} & U^{\top}
        \end{bmatrix} = 
        \Lambda^{\dagger}
        \begin{bmatrix}
            Z_{0}^{\top} & U^{\top}
        \end{bmatrix}
    \end{aligned}
\end{equation}

Substituting this simplified form back into Eq.~\eqref{eq_D3M} yields a final, explicit linear data-driven model:
\begin{equation}
    \begin{aligned}
        {y}_{k+1} & = Y\Lambda^{\dagger}
        \begin{bmatrix}
            Z_{0}^{\top} & U^{\top}
        \end{bmatrix}
        \begin{bmatrix}
            {z_{\mathrm{ini}}} \\
            {u}_k
        \end{bmatrix} \\
        & = \left(Y\Lambda^{\dagger}Z_{0}^{\top}\right){z_{\mathrm{ini}}} + \left(Y\Lambda^{\dagger}U^{\top}\right){u}_k
    \end{aligned}.
    \label{eq_D3M_simplified}
\end{equation}
This final expression is directly analogous to a discrete-time state-space representation, where the purpose of $z_{\mathrm{ini}}$ is similar to the initial states $x_{0}$, continuing the fitting stage of initial conditions. Then, applying any control inputs $u_k$ into the system under initial conditions, Eq. \eqref{eq_D3M_simplified} quantifies the corresponding system response via outputs $y_k$. Note that the system matrices $(Y\Lambda^{\dagger}Z_{0}^{\top})$ and $ (Y\Lambda^{\dagger}U^{\top})$ are computed purely from the collected data matrices $Z_0$, $U$, and $Y$. 


\subsection{Direct Data-driven Predictive Control}
A salient feature of the D3M framework is its ability to generate direct multi-step predictions via recursive application. Analogous to the traditional state-space representation, the D3M can be iteratively employed to forecast system outputs over a finite horizon of length $N_s$. Leveraging this property, we introduce a novel \emph{direct data-driven predictive control} (D3PC) scheme, formulated as follows. 

In particular, the prediction at step $k+2$ is obtained by propagating the system dynamics forward by one step, based on the available input and output trajectories. The one-step-ahead prediction for the output ${y}_{k+1}$ and the past-data vector ${z}_{k+1}$ (${z_{\mathrm{ini}}}$ at the $k+1$ step) are given by:
\begin{align}
    {y}_{k+1} &= (Y\Lambda^{\dagger}Z_{0}^{\top}){z}_{k} + (Y\Lambda^{\dagger}U^{\top}){u}_k, \\
    {z}_{k+1} &= (Z_1\Lambda^{\dagger}Z_{0}^{\top}){z}_k + (Z_1\Lambda^{\dagger}U^{\top}){u}_k,
\end{align}
where $Z_1$ is a data matrix constructed to represent the one-step shift in the past-data window, i.e., $Z_1=G\begin{bmatrix}
    Z_0^{\top}, U^{\top}, Y^{\top}
\end{bmatrix}^{\top}$; and $G$ is a rectangular forward-shift operator that removes the first block row.

To predict two steps ahead, we apply Eq. \eqref{eq_D3M_simplified} using the propagated state ${z}_{k+1}$ and the next control input ${u}_{k+1}$:
\begin{equation}
    \begin{aligned}
        {y}_{k+2} = (Y\Lambda^{\dagger}Z_{0}^{\top}){z}_{k+1} + (Y\Lambda^{\dagger}U^{\top}){u}_{k+1} 
    \end{aligned}.
\end{equation}
Substituting the expression for ${z}_{k+1}$ gives: 
\begin{equation}
    \begin{aligned}
        {y}_{k+2} & = (Y\Lambda^{\dagger}Z_{0}^{\top}) \Big( (Z_1\Lambda^{\dagger}Z_{0}^{\top}){z}_k + (Z_1\Lambda^{\dagger}U^{\top}){u}_k \Big) \\ 
        & + (Y\Lambda^{\dagger}U^{\top}){u}_{k+1} 
    \end{aligned}.
\end{equation}
Further, expanding the terms gives:
\begin{equation}
    \begin{aligned}
        {y}_{k+2} & = (Y \Lambda^{\dagger}Z_{0}^{\top} Z_{1} \Lambda^{\dagger} Z_{0}^{\top}){z}_{k} + (Y \Lambda^{\dagger} Z_{0}^{\top} Z_{1}\Lambda^{\dagger}U^{\top}){u}_{k} \\ 
        & + (Y\Lambda^{\dagger}U^{\top}){u}_{k+1}.
    \end{aligned}
\end{equation}
By continuing this recursive process for $N_s$ steps, we can represent the entire future output trajectory as a linear function of the initial past data ${z}_k$ and the sequence of future control inputs ${u}$. Then, we can establish a single, compact formulation that predicts the entire future output trajectory over a $N_s$-step horizon. This predictive model takes the form:
\begin{equation}
    {y} = \mathcal{M}{z_{\mathrm{ini}}} + \mathcal{S} {u},
    \label{eq_D3PC_prediction_model}
\end{equation}
where ${z_{\mathrm{ini}}} ={z}_k$ is the vector of initial past input-output sequence at step $k$, ${u} = [u_k^\top, \ldots, u_{k+N_s-1}^\top]^\top$ is the sequence of future control inputs, and ${y} = [y_{k+1}^\top, \ldots, y_{k+N_s}^\top]^\top$ is the predicted sequence of future outputs.

The system matrices $\mathcal{M}$ and $\mathcal{S}$ are constructed directly from the data matrices $U, Y, \Lambda, Z_0,$ and $Z_1$. To define these system matrices compactly, we first define the following intermediate matrices: 
\[
\begin{aligned}
    &\mathcal{M}_z = Y\Lambda^{\dagger}Z_{0}^{\top},\quad &&\mathcal{M}_u = Y\Lambda^{\dagger}U^{\top}, \\
    &\mathcal{A}_p = Z_1\Lambda^{\dagger}Z_{0}^{\top}, \quad &&\mathcal{B}_{p} = Z_1\Lambda^{\dagger}U^{\top}.
\end{aligned}
\]
Then, the matrix $\mathcal{M}$ is a tall matrix composed of block rows $\mathcal{M}_i$ for each step $i \in \{1, \dots, N_s\}$:
\begin{equation}
    \mathcal{M} = 
    \begin{bmatrix}
        \mathcal{M}_1 \\ \mathcal{M}_2 \\ \vdots \\ \mathcal{M}_{N_s}
    \end{bmatrix}, \quad \text{where} \quad
    \mathcal{M}_i = \mathcal{M}_{z}(\mathcal{A}_{p})^{i-1}
    \label{eq_M_matrix}
\end{equation}
The matrix $\mathcal{S}$ is a block lower-triangular matrix whose block element at the $i$-th row and $j$-th column is given by:
\begin{equation}
\mathcal{S}_{[i,j]} =
    \begin{cases}
        \mathcal{M}_{u} & \text{if } i = j \\
        \mathcal{M}_{z}(\mathcal{A}_{p})^{i-j-1}\mathcal{B}_{p} & \text{if } i > j \\
        0 & \text{if } i < j
    \end{cases}
    \label{eq_S_matrix}
\end{equation}
This compact representation is essential for the design of an efficient controller, as it circumvents the formatting difficulties associated with large-scale matrices such as $Z_0$, $U$, and $Y$, whose high column dimensions arise from the substantial data required to construct Hankel matrices. In other words, the dimensions of the matrices $\mathcal{M}$ and $\mathcal{S}$ are determined by the prediction horizon $N_s$ and the system orders instead of data volumes; for instance, $u_k \in \mathbb{R}^m$, $x_k \in \mathbb{R}^n$, and $y_k \in \mathbb{R}^p$ denote the dimensions of the inputs, states, and outputs, respectively.

With the predictive model established in Eq.~\eqref{eq_D3PC_prediction_model}, the D3PC is formulated as the following Quadratic Programming problem to be solved at each time step:
\begin{equation}
    \begin{aligned}
        \min_{{u}} \quad & \sum_{j=1}^{N_s}\left( \left\| y_{k+j} - r_{k+j} \right\|_{Q}^{2} + \left\| u_{k+j-1} \right\|_{R}^{2} \right) \\
        \mathrm{s.t.} \quad & {y} = \mathcal{M}{z_{\mathrm{ini}}} + \mathcal{S} {u} \\
        & {u} \in \mathcal{U}, {y} \in \mathcal{Y}
    \end{aligned}
    \label{eq_D3PM}
\end{equation}
where $r$ is the reference output trajectory, $Q$ and $R$ are weighting matrices, and $\mathcal{U}$ and $\mathcal{Y}$ are sets defining the input and output constraints, respectively.


\subsection{Implementation on Eco-driving Problem}
The implementation of the proposed D3PC for mixed traffic flow consists of three main steps: \textit{data collection}, \textit{D3M construction}, and \textit{formulation of the optimization problem}.  

At first, the acceleration signals for all CAVs are selected as control inputs, $u^d$. Then, velocity and inter-vehicle spacing information are selected as system outputs $(y^d = [v^d,s^d])$. With those collected trajectories, the input and output Hankel matrices for non-parametric representation of mixed traffic dynamics can be constructed. Then, the proposed D3M can be constructed via calculations in Eq. \eqref{eq_D3M_simplified}, indicating a data-driven model for mixed traffic flows.

Lastly, Eq. \eqref{eq_formulation} defines the eco-driving control objective for mixed traffic flows, which is to minimize overall energy consumption while maintain traffic safety and traffic flow efficiency. Thus, the D3PC control problem is formulated as a receding-horizon optimization that minimizes a weighted sum of predicted energy consumption, velocity deviations, and control efforts for the CAVs. The optimization is subject to constraints on the CAVs' acceleration limits and inter-vehicle spacing to ensure safety and traffic efficiency. 

\begin{equation}
    \label{eq_formulation}
    \begin{aligned}
        \min_{{u}} \quad & \sum_{k=0}^{N_s-1} \Big( \Vert y_{v,k} - r_{v,k} \Vert^2_Q + \Vert u_{k} \Vert^2_R + \Vert P_k \Vert^2_W \Big) \\
        \mathrm{s.t.} \quad & {y} = \mathcal{M}{z_{\mathrm{ini}}} + \mathcal{S} {u}, \\
        & u_{k} \in [a_{\mathrm{min}}, a_{\mathrm{max}}], \quad ~ i \in S  \\
        & y_{s,k} \in [s_{\mathrm{min}}, s_{\mathrm{max}}], \quad i \in S.
    \end{aligned}
\end{equation}
where $a_{\mathrm{min}}$, $a_{\mathrm{max}}$ are the acceleration bounds, $s_{\mathrm{min}}$ and $s_{\mathrm{max}}$ enforce safe and efficient inter-vehicle spacing, calculated by constant time headway and time gap \cite{li2025eco-driving}, and $P_k$ is the traction power calculated based on Eq.~\eqref{eq_polynomial}. Please refer to \cite{li2024physics} for more parameter specifications.

\section{RESULTS}
\subsection{Simulation Setup}
The simulation utilizes a 5-vehicle mixed traffic flow system ($S=\{1,3\}$) for verification. The test speed profile for the PV spans 6 minutes with a sampling rate of $0.1\mathrm{s}$, using the NGSIM dataset to represent real-world driving scenarios. Key control parameters are setting as follows, including the prediction horizon ($N_s=40$), past data length ($T_{\mathrm{ini}}=20$), future data length ($N=1$), and cost function weights ($Q=10, R=0.1, W=0.4$).

The primary focus of this study is to validate the controller's performance across a wide spectrum of diverse and unknown HDV behaviors. Two metrics related to eco-driving control performance are selected: energy consumption and computational time. To ensure a thorough and representative test of human driving diversity, we select a range of time headways from the distribution shown in Figure \ref{fig_Distribution}. The primary focus is on time headways in the range of $T_{\mathrm{gap}} \in [0.5,2.9]$s \cite{li2024physics}, divided into 12 equally spaced groups with a $0.2$s interval. Then, we randomly select two distinct human drivers (i.e., two random IDM headway parameters) from each group to create a test set that accurately reflects the majority of human driving behaviors. To represent almost all possible human driving behaviors, we consider all combinations of the test set, resulting in 576 scenarios. All simulations were conducted in MATLAB R2022b on a PC with a 3.8GHz processor and 64GB of RAM.

\subsection{Simulation Results}
\textbf{Computational efficiency Performance:}
A primary advantage of the D3PC method is its significant reduction in computational burden. To demonstrate the improvement in computational efficiency, the DeePC method used in \cite{li2024physics} is selected as the baseline. 

\begin{figure}[!htb]
    \centering
    \includegraphics[width=0.8\linewidth]{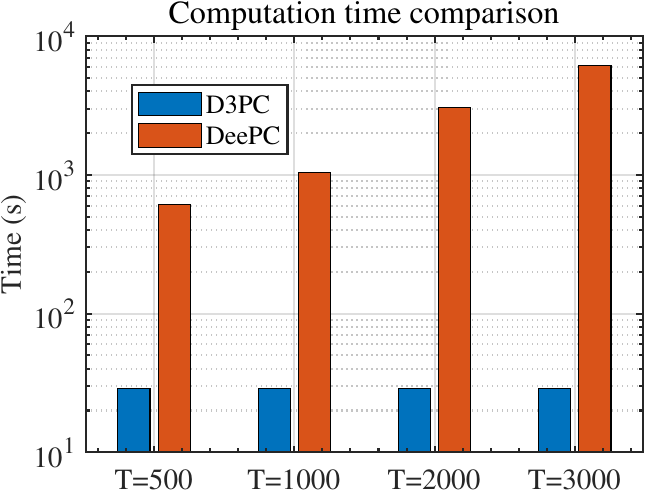}
    \caption{Computational time comparison between the D3PC and the DeePC.}
    \label{fig_computational_comparison}
\end{figure}

Figure \ref{fig_computational_comparison} compares the computation time of D3PC against the DeePC approach with different historical data lengths. The results are stark: the computation time for the DeePC grows substantially with the amount of data, quickly becoming prohibitive for real-time applications. In contrast, the D3PC's computation time is not only orders of magnitude lower but also remains nearly constant regardless of the data length. That is because the proposed D3PC method eliminates the computation of the high-dimensional vector (i.e., ${g}$ in DeePC), thereby simplifying the computational pathway. Thus, this remarkable efficiency makes the D3PC framework highly suitable for practical, real-time implementation on embedded vehicle hardware.

\textbf{Eco-driving Performance:}
The eco-driving performance is evaluated by examining the average energy consumption across all 576 simulation cases, each featuring a unique combination of diverse HDV behaviors. Table \ref{tab:EnergyConsumption_E_DeePC} provides a detailed comparison against three baselines: the DeePC controller with the same optimization problem and two OVM-based adaptive cruise control (ACC) controllers (an aggressive driver with $\alpha=0.8, \beta=0.5$ , one mild driver with $\alpha=0.5, \beta=0.8$) \cite{li2024physics}. 
\begin{table*}[!htp]
	\centering
	\caption{Platoon energy consumption under diverse driving characteristics.}
	\begin{tblr}{
	  cell{1}{2} = {c=7}{c},
        cell{2}{2-8} = {r=1}{c},
        cell{8}{1} = {c=8}{l},
        cell{9}{1} = {c=8}{l},
	  hline{1,3,8} = {-}{0.9pt},
	  hline{2} = {2-8}{},
	}
		  & Energy consumption (Average) (kJ) &               &                   &               &                  &               &                  \\
		  & D3PC  & DeePC & Improvement & OVM-1    & Improvement & OVM-2 & Improvement \\
	CAV1  & 1740.35 & 1865.22 & 6.69\%   & 1919.95 & 9.35\%   & 1886.48  & 7.75\%           \\
	HDV1  & 1715.44 & 1761.07 & 2.59\%   & 1895.27 & 9.49\%   & 1824.67  & 5.99\%           \\
	CAV2  & 1706.71 & 1778.07 & 4.01\%   & 1918.97 & 11.06\%  & 1814.87  & 5.96\%           \\
	HDV2  & 1704.13 & 1733.23 & 1.68\%   & 1956.10 & 12.88\%  & 1814.89  & 6.10\%           \\
	TOTAL & 6866.63 & 7137.59 & 3.80\%   & 7690.29 & 10.71\%  & 7340.91  & 6.46\%           \\
    Specifically, OVM-1 represents an aggressive human driver with $\alpha = 0.8 \text{ and } \beta = 0.5$, and OVM-2 \\ represents a mild human driver with $\alpha = 0.5\text{ and }\beta = 0.8$.
	\end{tblr}
	\label{tab:EnergyConsumption_E_DeePC}
\end{table*}

\begin{figure}[!htb]
    \centering
    \includegraphics[width=0.9\linewidth]{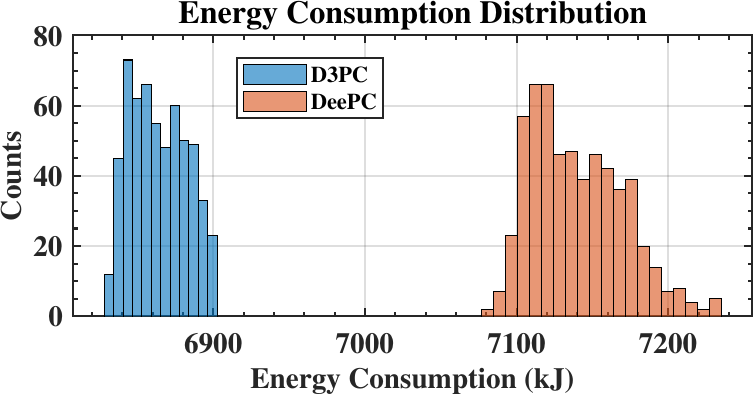}
    \caption{Energy consumption comparison between the D3PC and the DeePC.}
    \label{fig_EC_comparison_D3PC}
\end{figure}

The results highlight the superior efficiency of the D3PC. On average, D3PC achieves a total energy reduction of \textbf{3.80\%} compared to DeePC. The improvements over the OVM-based ACC controllers are even more substantial, with savings of \textbf{10.71\%} against the aggressive OVM-1 and \textbf{6.46\%} against the mild OVM-2. Figure \ref{fig_EC_comparison_D3PC} further visualizes these findings, showing that the energy consumption distribution for D3PC is clearly shifted to the left of DeePC, indicating consistently lower energy usage across the vast majority of test scenarios. This demonstrates the D3PC's ability to deliver robust and superior energy efficiency under a wide range of human driving behaviors.

\section{CONCLUSIONS}
This paper introduced a novel Direct Data-driven Predictive Control (D3PC) framework to make data-driven eco-driving computationally feasible for real-world mixed traffic flows. By reformulating the data-driven prediction mechanism of the conventional methods like DeePC and subspace predictive control, our method provides a computationally efficient predictive method whose complexity is nearly invariant to data size. 
This efficiency is critical for eco-driving, as it enables the real-time optimization needed to manage the unpredictable dynamics of human-driven vehicles while minimizing energy consumption. Extensive simulations validated that the D3PC is orders of magnitude faster than advanced DeePC baselines and delivers superior energy savings, reducing platoon consumption by up to 10.71\% compared to rule-based controllers. These findings confirm the D3PC's potential as a practical and effective solution for deploying safe and energy-efficient CAV control in complex mixed traffic environments.

\bibliographystyle{IEEEtran}
\bibliography{refs}

\end{document}